\documentclass[twocolumn,aps,pre,reprint,amsmath,amssymb]{revtex4-1}
\usepackage{graphicx}
\usepackage{dcolumn}
\usepackage{bm}
\usepackage{graphicx} 
\usepackage{epstopdf}
\usepackage{multirow}
\usepackage{CJK}
\usepackage{float}
\usepackage[colorlinks=true,dvipdfm]{hyperref}

\begin{document}


\title{Scaling Theories of Kosterlitz-Thouless Phase Transitions}

\author{Zhiyao Zuo, Shuai Yin, Xuanmin Cao, Fan Zhong}
\email{Corresponding author: stszf@mail.sysu.edu.cn}
\affiliation{State Key Laboratory of Optoelectronic Materials and Technologies, School of Physics, Sun Yat-Sen University, Guangzhou 510275, People's Republic of China}

\date{\today}

\begin{abstract}
We propose a series of scaling theories for Kosterlitz-Thouless (KT) phase transitions on the basis of the hallmark exponential growth of their correlation length. Finite-size scaling, finite-entanglement scaling, short-time critical dynamics, and finite-time scaling, as well as some of their interplay are considered. Relaxation times of both a normal power-law and an anomalous power-law with a logarithmic factor are studied. Finite-size and finite-entanglement scaling forms somehow similar to but different from a frequently employed ansatz are presented. The Kibble-Zurek scaling of topological defect density for a linear driving across the KT transition point is investigated in detail. An implicit equation for a rate exponent in the theory is derived and the exponent varies with the distance from the critical point and the driving rate consistent with relevant experiments. To verify the theories, we utilize the KT phase transition of a one-dimensional Bose-Hubbard model. The infinite time-evolving-block-decimation algorithm is employed to solve numerically the model for finite bond dimensions. Both a correlation length and an entanglement entropy in imaginary time and only the entanglement entropy in real-time driving are computed. Both the short-time critical dynamics in imaginary time and the finite-time scaling in real-time driving, both including the finite bond dimension, for the measured quantities are found to describe the numerical results quite well via surface collapses. The critical point is also estimated and confirmed to be $0.302(1)$ at the infinite bond dimension on the basis of the scaling theories.
\end{abstract}

\maketitle


\section{INTRODUCTION\label{intro}}
Kosterlitz-Thouless (KT) phase transitions are infinite-order phase transitions arising from binding of topological defect pairs for classical two-dimensional (2D) systems with a global $O(2)$ or $U(1)$ symmetry~\cite{Bere,KT73,Koster74}. In the low-temperature phase with a quasi-long-range order, the correlation length diverges and correlations decay algebraically at large distance with an exponent varying continuously with temperature. No true long-range order and hence nonvanishing order parameter appear, in accordance with the Mermin-Wegner theorem~\cite{MW,Hohenberg}. The high-temperature disordered phase consists of unbound defects and has usual exponentially decaying correlations with an exponentially rather than the usual algebraically increasing correlation length $\xi$ upon approaching the critical point, viz.~\cite{Koster74},
\begin{equation}
\xi=\xi_0\exp\left(ag^{-\nu}\right),
\label{eq1}
\end{equation}
where $\xi_0$ and $a$ are nonuniversal constants, $g$ measures the distance to the critical point, and $\nu=1/2$. Because of the essential singularity at the critical point in Eq.~(\ref{eq1}), the free energy and all its derivatives are continuous. The rapidly increasing correlation length renders numerical and experimental determination of critical properties challenging. This is worsened by logarithmic corrections~\cite{Weber,Wallin,Mishra,Pino,Carra,Dalmonte,Butera93,Kenna95,Irving,Campo,Janke,Kenna,Jaster,Balog,Balogc,Chandra,Hasenbusch05,Butera,Arisue,Komura,Hsieh,Keesman,Diaz,Bray,Ying,Zheng,Lei} to asymptotic behavior owing to a marginal operator at the fixed point predicted by perturbation renormalization-group (RG) calculations~\cite{Koster74,Jose,Amit,Balog,Pelissetto}. Even subleading logarithmic corrections were found to contribute appreciably~\cite{Hasenbusch05,Hsieh}.

The KT phase transition is found in a lot of systems~\cite{Karimov,vanBei,Bishop,Beasley,Hebard,Bishop80,Resnick,Nienhuis,Kawabat,Pargellis,Elmers,Romano,Caqpriotti,Maggiore,Weigel,Gasparini,Hadz06,Kruger,Hadz,Taroni,Clade,Hung,Giamarchi,Dalmonte}. Here we focus on the one in the 1D
Bose-Hubbard (BH) model~\cite{Gersch,fisher,Krutitsky}, which can be studied experimentally in optical lattices~\cite{Stoferle,Giamarchi04,Caza}. The model consists essentially of a kinetic energy and a repulsive on-site interaction whose competition leads to quantum phase transitions~\cite{Sachdev} from a Mott insulating phase to a superfluid phase as the relative strength of the interaction decreases. The transition can be induced either by density fluctuations or by phase fluctuations at constant density. Whereas the first kind of the transitions is governed by a Gaussian fixed point, the second one is a KT transition and occurs only at commensurate fillings. Thus, the gapped Mott insulator phase has its gap asymptotically proportional to $\xi^{-z}$ with $\xi$ obeying Eq.~(\ref{eq1}) and hence the gap closes exponentially as the transition point is approached, while the gapless superfluid phase possesses only quasi-long range order, where $z$ is the dynamic critical exponent. The exponentially decreasing energy gap again renders its study difficult~\cite{Krutitsky}.

The Kibble-Zurek (KZ) mechanism for topological defect formation during a cooling through a usual critical point from a disordered phase to an ordered phase~\cite{KZ1,Kibble2,KZ2,KZ3,Dziarmaga,Polkovnikov,inexper4} has been extended to the KT universality class in the 1D BH model~\cite{Bernier,Braun,Dzia,Gardas,Weiss} as well as other models~\cite{Jelic,Deutsch}. In contrast to the usual KZ scaling for the dependence of the defect density on the cooling rate with a fixed exponent, the exponent found experimentally varies with the driving rate and the distance away from the critical point in the 1D BH model~\cite{Braun}. Effective exponents were proposed and tested to approximate the real behavior of the KT transition in the 1D BH model with power laws within sufficiently small ranges~\cite{Dzia,Gardas,Weiss}. Scaling using the correlation length has been investigated in general~\cite{fisher,Pelissetto} and tested for the 1D BH model with driving~\cite{Gardas}. A different scaling form considering the effect of phase ordering below the critical temperature was also suggested and examined~\cite{Jelic}.

In general driven nonequilibrium critical phenomena~\cite{Feng} arising from driving with finite rates through a usual critical point, a theory of finite-time scaling (FTS)~\cite{Zhong1,Zhong2,Huang} has been proposed on the basis of a dynamic RG theory~\cite{Zhong06}. This is obtained by realizing that the driving rate $R$ contributes a finite timescale $R^{-z/r}$ that plays a role similar to what the lattice size plays in the famous finite-size scaling, where $r$ denotes the rate exponent which is the RG eigenvalue of $R$ and is a constant for a usual critical point. On the other hand, for the nonequilibrium ``initial slip" of a system quenched rapidly from a nonequilibrium initial condition to near its critical point~\cite{Janssen}, a short-time critical dynamics method has been widely applied to many systems~\cite{Li,Zheng96,Zheng99,Albano}, including the KT phase transitions in classical systems~\cite{ZhengXY,Ying,Zheng,Lei}. A quantum version of short-time critical dynamics for imaginary time has also been developed and successfully applied to both usual and topological quantum phase transitions~\cite{Yin,Zhang}.

Here, we propose a series of scaling theories for the KT phase transitions built on its hallmark exponential growth of the correlation length $\xi$ in Eq.~(\ref{eq1}). The salient feature of the theories is to employ directly the distance to the critical $g$ as a variable or scaling field, instead of using $\xi$ itself to scale time and space as usual~\cite{fisher,Pelissetto,Gardas,Ozeki,Ozeki20}. This is because $g$ is directly controllable whereas $\xi$ has to be computed from a given $g$. In order to verify the theories, we study the dynamics of the 1D BH model using the infinite time-evolving-block-decimation (iTEBD) algorithm~\cite{Vidal,Vidal03,Vidal04,Vidalitebd}. This necessitates consideration of the bond dimension $D$ of the number of states kept owing to the exponential growth of $\xi$. For an imaginary-time evolution, the scaling theory of universal short-time quantum critical dynamics for a finite $D$ is then required. For a real-time linear driving, the FTS theory with again a finite $D$ is needed. The most prominent result is that the rate exponent $r$ is now both scale and $g$ dependence and is not a constant anymore. Both the scaling theories agree with the numerical results reasonably well. Meanwhile, we determine and confirm the KT transition point in the model utilizing the short-time critical dynamics and the fact that the superfluid phase has $z=1$.

We emphasize that most results obtained are new. The most important one is the scaling theories in terms of the scaling field $g$, including both the short-time critical dynamics and FTS testified for the 1D BH model and those not yet tested. These theories enable us to systematically consider the effects of multiple scales. The explicit expressions for the rate exponent for both kinds of relaxation time to be specified later on, the short-time critical dynamics in imaginary time itself and its scaling collapses in the presence of the finite $D$ and similar collapses for FTS of the 1D BH model have not been investigated before. Even the method to determine the KT transition point is new. The scaling directly with $g$ facilitates its application even though the perturbative RG theory for the KT transitions has been well developed; and the successful verification the scaling theories in the 1D BH model indicates that they can be extended to other KT transitions as well.

In the following, we first develop the scaling theories of the KT phase transitions in Sec.~\ref{scaling}. Then, after introducing the BH model and the method in Sec.~\ref{model}, we present in Sec.~\ref{num} the numerical results of the 1D BH model using the iTEBD algorithm to verify the theories. In Sec.~\ref{stcd}, we apply the short-time critical dynamics with finite bond dimensions to examine the imaginary-time evolution of the correlation length and the entanglement entropy and to determine the KT phase transition point; while in Sec.~\ref{ftswd} we focus on the FTS with finite bond dimensions to study the real-time driving dynamics from the Mott phase to the superfluid phase. Finally, we summarize the results in Sec.~\ref{summ}.

\section{Scaling theories\label{scaling}}
In this section, we present our scaling theories for the KT phase transitions. As emphasized, we employ directly the distance to the critical point $g$ as a variable. In the following, we first derive how $g$ transforms under a length rescaling in Sec.~\ref{stog} and, on the basis of it, we successively study finite-size scaling, short-time critical dynamics, and finite-time scaling in Secs.~\ref{fss},~\ref{tstcd}, and~\ref{tfts}, respectively. Finite-entanglement scaling is briefly mentioned in Sec.~\ref{tstcd}. In Secs.~\ref{tstcd} and~\ref{tfts}, two different forms of relaxation time are considered.

\subsection{\label{stog}Scale transformation of $g$}
In order to see how the scaling theories are derived, we start with a usual critical point in the proximity of which the correlation length as a function of $g$ diverges as
\begin{equation}
\xi(g)\sim g^{-\nu},\label{xius}
\end{equation}
asymptotically with the correlation length critical exponent $\nu$, in contrast with Eq.~(\ref{eq1}) for the KT transition. Here as mentioned above, $g$ measures the distance to the critical point, which can be a reduced temperature $|T-T_c|$ for a thermal phase transition or a reduced coupling $|J-J_c|$ for a quantum phase transition, where $T$ is the temperature, $J$ is a parameter in a Hamiltonian, and the subscripts denote their corresponding critical values. As scales are changed by a rescaling factor $b$ in a renormalization,
\begin{equation}
\xi(g(b))=\xi(g)/b,\label{xib}
\end{equation}
viz., the correlation length is reduced by the factor $b$. Naturally, the algebraic divergence of the correlation length with $g$ holds for $\xi(g(b))$ provided that it falls within the critical region, i.e.,
\begin{equation}
\xi(g(b))\sim g(b)^{-\nu}.\label{xiusg}
\end{equation}
Plugging Eqs.~(\ref{xiusg}) and~(\ref{xius}) into Eq.~(\ref{xib}) results in
\begin{equation}
g(b)=gb^{1/\nu},\label{gbus}
\end{equation}
which is just the usual scale transformation of $g$. In a scaling theory, one reverses the above reasoning and starts with Eqs.~(\ref{xib}) with~(\ref{gbus}) as an ansatz. Choosing $b=g^{-\nu}$ then leads back to Eq.~(\ref{xius}).

Now, for the KT phase transitions with their hallmark correlation length diverging as Eq.~(\ref{eq1}), assuming again that $g(b)$ is not too far away from the critical point so that the correlation length at $g(b)$, $\xi(g(b))$, obeys the same exponentially diverging form, viz.,
\begin{equation}
\xi(g(b))=\xi_0\exp\left(ag(b)^{-\nu}\right)
\label{xig}
\end{equation}
in place of Eq.~(\ref{xiusg}), we then have
\begin{equation}
 g(b)=g\left(1-\frac{g^\nu}{a}\ln b\right)^{-1/\nu}
 \label{eq2}
\end{equation}
in place of Eq.~(\ref{gbus}) following the same reasoning. One can employ Eq.~(\ref{eq1}) to rewrite Eq.~(\ref{eq2}) in another form
\begin{equation}
 g(b)=g\left[1-\frac{\ln b}{\ln(\xi/\xi_0)}\right]^{-1/\nu},
 \label{eq2x}
\end{equation}
which will be invoked to make contact with some frequently quoted approximations. Using Eqs.~(\ref{eq1}) and~(\ref{xib}), one can convince oneself that Eqs.~(\ref{eq2}) and~(\ref{eq2x}) are in fact $g(b)^{-\nu}/g^{-\nu}=\ln (\xi(g(b))/\xi_0)/\ln(\xi/\xi_0)$, which is true due to Eqs.~(\ref{eq1}) and~(\ref{xig}). At the critical point, $g(b)=g=0$.

Equation~(\ref{eq2}) dictates the scale transformation of the distance to the critical point as scales are changed. One sees that it is now not a simple power law of $b$ like Eq.~(\ref{gbus}) due to the exponential divergence. As a result, the usual classification of the relevancy of an operator according to its renormalization-group eigenvalue does not work. Since the dependence of the rescaling factor is logarithmic, we may loosely refer $g$ as logarithmic relevance.

Given Eq.~(\ref{eq2}) or~(\ref{eq2x}), in a scaling theory for the KT transitions based on Eq.~(\ref{xib}) with $g$ as the only variable, choosing $b=\xi_0\exp(ag^{-\nu})/\xi(C)$ such that $g(b)=C$, a constant as usual, then results back in Eq.~(\ref{eq1}) consistently, where $\xi(C)=\xi_0\exp(aC^{-\nu})$, the correlation length at $C$.

In the perturbative RG theory for the KT phase transitions~\cite{Koster74,Jose,Amit,Balog,Pelissetto}, there are two RG equations, one for an effective temperature and the other an effective chemical potential controlling the density of defects. However, both variables serve as coupling constants and the scale dependence of the reduced temperature is seldom considered. The reduced temperature $|T_c-T|$ appearing in Eq.~(\ref{eq1}) in place of $g$ comes from the initial condition for the two couplings. Moreover, upon combining the two equations, an RG invariant function can be found which is assumed to play the role of the thermal scaling field and is expanded as a series of the reduced temperature~\cite{Pelissetto}. However, we emphasize that, once Eq.~(\ref{eq1}) is valid, the variation of the correlation length $\xi$ with the scale must lead to Eq.~(\ref{eq2}), the scale transformation of $g$. In fact, in Eq.~(\ref{eq2}), the second term within the parentheses is small near the critical point because of the small $g$ and finite $a$ and $b$. Therefore, one sees that to the lowest order in $g$, the reduced temperature $g$ is indeed invariant under scale transformations. The higher-order corrections depend only on $\ln b$, the logarithm of the scale. Probably, this logarithm is the origin of the logarithmic corrections in the KT phase transitions. Note that the RG theory is only perturbative, while Eq.~(\ref{eq2}) may hopefully catch non-perturbative effects optimistically.
On the other hand, if one abides by the perturbative RG results, higher order terms contribute to Eq.~(\ref{eq1}) as an analytic series of $g$ as~\cite{Pelissetto}
\begin{equation}
 \xi=\xi_0\exp\left[ag^{-\nu}(1+a_1g+\cdots)\right],
\label{eq1c}
\end{equation}
where $a_1$ is a nonuniversal constant. One can then derive perturbatively corrections to Eq.~(\ref{eq2}) using the same method straightforwardly. Here, we are content with the leading result expressed in Eq.~(\ref{eq2}), since many practical applications of scaling theories consider only the leading behavior. We will show that the leading behavior can still describe well the scaling near the KT point within some finite regions when other scaling fields are taken properly into account.

\subsection{\label{fss}Finite-size scaling}
To verify Eq.~(\ref{eq2}), one has to consider other variables, since, from Eq.~(\ref{eq1}), for $a=2$ and $\nu=1/2$, a $g=10^{-2}$ gives rise to a huge $\xi/\xi_0\approx4.8\times10^8$, which may well exceed other length scales. A frequently considered variable is the lateral size $L$ of a finite system. This leads to the well-known finite-size scaling. Although we will not need the finite-size scaling in analysing our numerical results, we still present it here as a form of the scaling theories for the sake of completeness. This is justified since finite-size scaling is frequently employed in the KT transitions~\cite{Weber,Wallin,Mishra,Carra,Dalmonte}.

For a system with a finite size $L$, one needs to add $L^{-1}b$ to Eq.~(\ref{xib}), viz.,
\begin{equation}
\xi(g(b),L^{-1}b)=\xi(g,L^{-1})/b,\label{xibL}
\end{equation}
and obtains
\begin{equation}
\xi(g,L^{-1})=Lf_L(g(1-g^{\nu}\ln L/a)^{-1/\nu})
\label{xiL}
\end{equation}
upon choosing $b=L$, or, using Eqs.~(\ref{eq2x}) and~(\ref{xiL}),
\begin{equation}
\xi(g,L^{-1})=Lf_L(g\{\ln(f_L/\xi_0)/[\ln L+\ln(f_L/\xi_0)]\}^{-1/\nu}),
\label{xiLx}
\end{equation}
where $f_L$ is a universal scaling function. Equations~(\ref{xiL}) and~(\ref{xiLx}) are finite-size scaling forms for the KT transitions. Defining another scaling function $\tilde{f}_L(y)=f_L(x)=\xi/L$ with a new argument $y\equiv ax^{-\nu}$, one finds,
\begin{eqnarray}
y&=&\tilde{f}_L^{-1}(\xi/L)= \ln(\xi/\xi_0)\frac{\ln(f_L/\xi_0)}{\ln L+\ln(f_L/\xi_0)}\nonumber\\
&\simeq&\ln(\xi/\xi_0)\ln\left[1+\frac{\ln(f_L/\xi_0)}{\ln L+\ln(f_L/\xi_0)}\right]
\label{xiLxp}
\end{eqnarray}
using Eq.~(\ref{eq1}) and $x$ in Eq.~(\ref{xiLx}), where $\tilde{f}_L^{-1}$ is the inverse function of $\tilde{f}_L$ and the approximation is good for large $L$. Equation~(\ref{xiLxp}) is somehow similar to an ansatz~\cite{Weber,Wallin,Mishra,Carra,Dalmonte}
\begin{equation}
L\Delta(L)\left(1+\frac{1}{2\ln L+C_0}\right)=F(\xi/L),
\label{ansatz}
\end{equation}
proposed to estimate $T_c$, where $C_0$ is a constant, $F$ is a scaling function, and $\Delta(L)$ denotes the energy gap at size $L$. However, they are different, even if we take the corrections in Eq.~(\ref{eq1c}) into account. One can multiply the whole denominator $\ln L+\ln(f_L/\xi_0)$ in the second line of Eq.~(\ref{xiLxp}) by $2$. Then $y$ becomes $y^{\nu}$ and the first factor, $\ln(\xi/\xi_0)$, changes to a more complicated function of $\xi$, viz., $\ln^{\nu}(\xi/\xi_0)$. This explains the dependence of the number $2$ in Eq.~(\ref{ansatz}) on the employed variable~\cite{Dalmonte}. Note that $f_L$, as defined in Eq.~(\ref{xiL}), depends on $g$ and $L$ and thus Eq.~(\ref{xiLx}) is as exact as Eq.~(\ref{xiL}). However, for small $x$, one can approximate $f_L(x)\simeq f_L(0)$, a constant. In this case, according to Eq.~(\ref{xiLxp}), one may adjust the critical point, $\xi_0$, $a$, and the constant $\ln(f_L(0)/\xi_0)$ such that the measured $y$ versus $\xi/L$ for various $g$ and $L$ collapses approximately onto a single curve given by $\tilde{f}_L^{-1}$. To overcome the approximation, one can employ Eq.~(\ref{xiL}) and plots the measured $\xi/L$ versus $g(1-g^{\nu}\ln L/a)^{-1/\nu}$ by selecting the critical point and $a$, now only two parameters, such that all data collapse onto a single curve representing $f_L$. Note that $\nu=1/2$ is an input in all these treatments. Note also that $\xi_0\exp[ag(1-g^{\nu}\ln L/a)^{-1/\nu}]^{-\nu}$ is just $\xi/L$ and Eq.~(\ref{xiL}) has indeed the correct form. However, as mentioned, using $g$ instead of $\xi$ is more convenient to analyze data and to derive a series of unique results in this paper. Equations~(\ref{xiL}),~(\ref{xiLx}), and~(\ref{xiLxp}) are open for test; here we employ iTEBD for an infinite chain~\cite{Vidalitebd} and do not need to consider finite-size scaling.

\subsection{\label{tstcd}Short-time critical dynamics}
Starting from this section, we consider the time evolution of the KT transitions. We will study first the short-time critical dynamics~\cite{Zheng99} in this section and leave the FTS to the next section. These necessitate the introduction of the time $t$. It may either be the imaginary time or be the real time, though the so-called ``initial slip" critical exponent~\cite{Janssen} was found to be slightly different for the imaginary-time evolution of the quantum phase transition in the transverse-field Ising model and the real-time evolution of the classical Ising model near its critical point~\cite{Yin}. Since we employ TEBD to study the time evolution, we have to take into account simultaneously the bond dimension $D$~\cite{entangle1,Cao}, which is the number of states kept in matrix product states~\cite{Klumper,Verstraete1,Schollwock}. This is related to finite-entanglement scaling~\cite{entangle1,Cao}. Moreover, we will consider two kinds of relaxation time. One is the normal one in which the relaxation time is a simple power law of the correlation length and the other anomalous one contains an additional logarithmic corrections. The latter is relevant for KT transitions with logarithmic defect mobility such as the $XY$ model and liquid-crystal systems~\cite{Dubois,Ryskin,Rad}.

With $t$ and $D$, Eq.~(\ref{xib}) is extended to be
\begin{equation}
\xi(g,t,D^{-1})=b\xi(g(1-g^{\nu}\ln b/a)^{-1/\nu},tb^{-z},D^{-1} b^{1/\kappa}),
\label{xibtx}
\end{equation}
where $\kappa$ is a universal exponent. It was found analytically to be related to the central charge $c$ through
\begin{equation}
\kappa=\frac{6}{c\left(\sqrt{12/c}+1\right)}
\label{kappa}
\end{equation}
rather than to the scaling dimension of an operator~\cite{kappav}. In Eq.~(\ref{xibtx}), we have written $D^{-1}$ for $D$ similar to the appearance of $L$ as $L^{-1}$ in Sec.~\ref{fss} in consideration of the fact that no dependence on it appears when it is infinity. Similarly, we should have also written $t$ as $t^{-1}$. However, we will not consider explicitly scaling functions that contain $t$ as their scaled variables except for Eq.~(\ref{xitpx}) below. Consequently, there exists no infinity for equilibrium at infinite time. In addition, we do not take into account the scale dependence of an initial finite order parameter. This is because we only consider nonequilibrium relaxation from a highly ordered initial state whose large order parameter renders the initial slip region rather short in time~\cite{Li,Zheng96,Zheng99,Yin,Zhang}.

From Eq.~(\ref{xibtx}), setting $b=t^{1/z}$ results in the short-time critical dynamic scaling form
\begin{equation}
\xi(g,t,D^{-1})=t^{1/z}f_{t}(g[1-g^{\nu}\ln t/(az)]^{-1/\nu},D^{-1} t^{1/z\kappa}),
\label{eq:four}
\end{equation}
where $f_{t}$ is another scaling function. A form similar to Eq.~(\ref{xiLx}) can also be derived. On the other hand, if $b=D^{\kappa}$, one finds
\begin{equation}
\xi(g,t,D^{-1})=D^{\kappa}f_{D}(g(1-\kappa g^{\nu}\ln D/a)^{-1/\nu},tD^{-z\kappa}),
\label{fes}
\end{equation}
with yet another scaling function $f_D$ for finite-entanglement scaling because of the finite entanglement limited by the bond dimension~\cite{entangle1,Cao}. Accordingly, $\xi\sim D^{\kappa}$ asymptotically correctly. One can also replace $L$ and $f_L$ in Eqs.~(\ref{xiLx}) and~(\ref{xiLxp}) with $D^{\kappa}$ and $f_D$, respectively, and cast the finite-entanglement scaling in a form similar to the ansatz~(\ref{ansatz}) in finite-size scaling.

We will utilize Eq.~(\ref{eq:four}) to determine the critical properties of the KT transition. To this end, we take the logarithm of Eq. (\ref{eq:four}) to obtain
\begin{equation}
	\ln \xi=\frac{1}{z}\ln t+\ln f_{t}(g[1-g^{\nu}\ln t/(az)]^{-1/\nu},D^{-1} t^{1/z\kappa}).
	\label{eq4}
\end{equation}
Accordingly, exactly at the critical point $g=0$ for $D=\infty$, the slope of $\ln\xi$ versus $\ln t$ is just $1/z$. However, for a finite $D$, the slope depends on $t$. Yet, above the critical point where Eq.~(\ref{eq1}) is valid, the $1/z$ slope returns for fixed $g[1-g^{\nu}\ln t/(az)]^{-1/\nu}$ and $D^{-1} t^{1/z\kappa}$ such that a fixed $D$ correspond to a fixed $t$ and $g$. Below the critical point, the correlation length diverges and Eq.~(\ref{eq4}) derived from Eq.~(\ref{eq1}) fails. However, $\xi$ obtained from a finite $D$ may well be finite owing to the finite-entanglement scaling~\cite{Cao}.

We will corroborate the theory using the entanglement entropy, which is related to the correlation length through~\cite{Cala}
\begin{equation}
S\approx \frac{c}{6}\ln\xi\label{scx}
\end{equation}
with the central charge $c$. Therefore, its short-time critical dynamic scaling form is
\begin{equation}
S\approx\frac{c}{6z}\ln t+\frac{c}{6}\ln f_{t}(g[1-g^{\nu}\ln t/(az)]^{-1/\nu},D^{-1} t^{1/z\kappa}).
\label{eq:iment}
\end{equation}

Now we consider the case with an anomalous relaxation time. Equation~(\ref{eq:four}) yields $\xi\sim t^{1/z}$ at the critical point $g=0$ for $D=\infty$, as expected for usual quantum and classical critical points~\cite{Janssen,Yin,Zhang}. However, for the 2D $XY$ model that is a generic example of the KT transitions, it was found that this is only true for an ordered initial condition; whereas $\xi\sim [t/\ln (t/t_0)]^{1/z}$ for a disordered initial condition, both at and below the critical point, where $t_0$ is a microscopic time scale~\cite{Bray}. However, the logarithmic factor was found to be absent for a quench from well above $T_c$ to a low temperature in a 2D superfluid~\cite{Forrester}. In fact, such a logarithmic factor prevails in phase-ordering kinetics~\cite{Bray94} of the $XY$ model at low temperatures~\cite{Toyoki,Mondello,Yurke,Rutenberg,Jelic} and should arise from the logarithmic mobility of topological defects in systems such as the 2D $XY$ model and liquid crystals~\cite{Dubois,Ryskin,Rad}. In addition, $z=2$ rather than the usual dynamics critical exponent because of the diffusive nature of the growth~\cite{Toyoki,Mondello,Yurke,Rutenberg}. As such, similar behavior ought to appear above the critical point. Consequently, the relaxation time for the correlation length behaves as~\cite{Jelic}
\begin{equation}
\tau_{\xi}\simeq\xi^z\ln(\xi/\xi_0),\label{teq}
\end{equation}
rather than the usual
\begin{equation}
\tau \sim \xi^z.\label{txiz}
\end{equation}
To account for this unusual behavior, we can replace the usual $t(b)=tb^{-z}$ with
\begin{equation}
t(b)=tb^{-z}\left(1-\frac{g^\nu}{a}\ln b\right),
\label{tb}
\end{equation}
in Eq.~(\ref{xibtx}). Indeed, choosing again $b\propto\exp(ag^{-\nu})$ such that $g(b)$ of Eq.~(\ref{eq2}) is a constant then leads to
\begin{equation}
\xi=\exp\left(ag^{-\nu}\right)f_{\tau}(tg^{\nu}\exp(-zag^{\nu}),D^{-1} \exp(ag^{-\nu}/\kappa))
\label{xitpx}
\end{equation}
with a new scaling function $f_{\tau}$. This is a quasiequilibrium scaling form for $\xi$ at sufficiently long times and large $D$ such that the two scaled variables in $f_{\tau}$ are negligible. According to Eq.~(\ref{xitpx}), a characteristic time scale is given by $tg^{\nu}\exp(-azg^{\nu})={\rm constant}$, which reproduces Eq.~(\ref{teq}) correctly. On the other hand, setting $t(b)=1/C_1$, a constant, one finds $b\simeq\{-C_1t[1-g^{\nu}\ln (C_1t)/(az)]\}^{1/z}$ and hence the evolution of the correlation length follows
\begin{eqnarray}
\xi=\left[C_1t\left(1-\frac{\ln (C_1t)}{z a g^{-\nu}}\right)\right]^{\frac{1}{z}}\!\!\bar{f}_t(g[1-g^{\nu}\ln (C_1t)/(az)]^{-1/\nu},\nonumber\\
D^{-1} (C_1t)^{1/z\kappa}[1-g^{\nu}\ln(C_1t)/(az)]^{1/z\kappa}),\qquad\qquad\quad
\label{xitppx}
\end{eqnarray}
upon neglecting terms of order $\ln\ln t$ and higher, where $\bar{f}_t$ is yet another scaling function. Equation~(\ref{xitppx}) seems to give $\xi\sim t^{1/z}$ at the critical point $g=0$ again. However, we can employ Eq.~(\ref{eq2x}), which involves $\ln (\xi/\xi_0)$ that evolves with time. One then arrives at an exact relation
\begin{equation}
b=\left\{\frac{C_1t\ln(\bar{f}_t/\xi_0)^z}{\ln\left[C_1t(\bar{f}_t/\xi_0)^z\frac{\ln (\bar{f}_t/\xi_0)}{\ln (\xi/\xi_0)}\right]}\right\}^{1/z}
\label{bt}
\end{equation}
from Eq.~(\ref{tb}) and the definitions of $C_1$ and $\bar{f}_t$ in Eq.~(\ref{xitppx}). Upon neglecting terms of order $\ln\ln (t/t_0)$ and higher with $t_0^{-1}=C_1(\bar{f}_t/\xi_0)^z\simeq C_1(\bar{f}_t(0,0)/\xi_0)^z$, Eq.~(\ref{bt}) becomes $b\simeq[-C_1t\ln (C_1t_0)/\ln(t/t_0)]^{1/z}$ and the correlation length now grows as
\begin{eqnarray}
\xi=\left[\frac{-C_1t\ln (C_1t_0)}{\ln (t/t_0)}\right]^{1/z}\bar{f}_t(g[-\ln (C_1t_0)/\ln(t/t_0)]^{-1/\nu},\nonumber\\
D^{-1}[-C_1t\ln (C_1t_0)/\ln(t/t_0)]^{1/z\kappa}),\qquad\qquad\qquad\quad\label{xitppx1}
\end{eqnarray}
which indeed exhibits the growth of $[t/\ln(t/t_0)]^{1/z}$ consistently.

Whether the special relation time $\tau_{\xi}$ and the resultant scaling theories apply to the quantum KT transitions or not and whether the classical diffusive growth of the correlation length is relevant to the quantum cases or not are interesting open questions to be investigated.

\subsection{\label{tfts}FTS}
We now study the driving nonequilibrium critical phenomena via changing $g$ linearly with the real rather than imaginary time, viz.,
\begin{equation}
g=Rt,\label{grt}
\end{equation}
with the constant rate $R$. We have chosen the time origin at the critical point. Accordingly, a driving with positive $R$ starts at some initial time $t<0$ in the disordered phase, passes through $t=0$ at the critical point and ends in the ordered phase. This introduces $R$ into the scaling hypothesis Eq.~(\ref{xibtx}) with $t$ being the real time now. However, because $g$, $R$, and $t$ are related by Eq.~(\ref{grt}), only two out of the three are independent. We choose $g$ and $R$ as independent variables. As a result, in place of Eq.~(\ref{xibtx}), we have now
\begin{equation}
\xi(g,D^{-1},R,b)=b\xi(g(1-g^{\nu}\ln b/a)^{-1/\nu},D^{-1} b^{1/\kappa},Rb^{r}).
\label{eq3}
\end{equation}
The RG eigenvalue $r$ of $R$ is determined by assuming Eq.~(\ref{grt}) is valid in the scale $b$, i.e., $g(b)=R(b)t(b)$ as usual~\cite{Zhong1,Zhong2}. This results in
\begin{equation}
b^{r-z}=\left(1-\frac{g^\nu}{a}\ln b\right)^{-1/\nu},
\label{brz}
\end{equation}
which is now both scale $b$ and $g$ dependent instead of the usual scaling law~\cite{Zhong1,Zhong2}
\begin{equation}
r=z+1/\nu,\label{rznu}
\end{equation}
obtained if the righthand side of Eq.~(\ref{brz}) is just $b^{1/\nu}$ from Eq.~(\ref{gbus}) instead of Eq.~(\ref{eq2}). From Eq.~(\ref{eq3}), choosing $b=R^{-1/r}$ leads to the FTS form
\begin{equation}
\xi=R^{-1/r}f_{R}(g[1+g^{\nu}\ln R/(ar)]^{-1/\nu},D^{-1} R^{-1/r\kappa}),
\label{eq:fts}
\end{equation}
with a scaling function $f_R$, where $r$ is now given implicitly by
\begin{equation}
R^{z/r-1}=\left(1+\frac{g^\nu}{ar}\ln R\right)^{-1/\nu}
\label{eq6}
\end{equation}
from Eq.~(\ref{brz}) self-consistently. Equation~(\ref{eq6}) allows us to write Eq.~(\ref{eq:fts}) in a simplified form,
\begin{equation}
\xi(g,R,D^{-1})=R^{-1/r}f_{R}(gR^{-(r-z)/r},D^{-1} R^{-1/r\kappa}),
\label{eq:ftsr}
\end{equation}
which is just the usual FTS form $gR^{-1/r\nu}$ if Eq.~(\ref{rznu}) is valid~\cite{Zhong1,Zhong2}.
Similar to Eq.~(\ref{eq:iment}), the FTS form of the entanglement entropy is
\begin{equation}
	S(g,R,D^{-1})\approx-\frac{c}{6r}\ln R+\frac{c}{6}f_{R}(gR^{-(r-z)/r},D R^{1/r\kappa}).
	\label{eq:rs}
\end{equation}

For the sake of completeness, we also present the results for $t$ transforming according to Eq.~(\ref{tb}) for the anomalous relaxation time Eq.~(\ref{teq}). In this case, upon assuming again $R(b)=Rb^r$, Eqs.~(\ref{eq2}),~(\ref{tb}), and~(\ref{grt}) result in an equation similar to Eq.~(\ref{brz}) with $-1/\nu-1$ in place of the exponent $-1/\nu$. After setting $b=R^{-1/r}$, it becomes
\begin{equation}
R^{z/r-1}=\left(1+\frac{g^\nu}{ar}\ln R\right)^{-1/\nu-1},
\label{brz1}
\end{equation}
in place of Eq.~(\ref{eq6}). However, the FTS forms for $\xi$ and $S$, Eqs.~(\ref{eq:fts}),~(\ref{eq:ftsr}), and~(\ref{eq:rs}), remain intact if the same pair of independent variables is picked since the effect of the time transformation has embodied in $r$. On the other hand, if we select $t$ and $R$ as independent variables, the FTS scaling form of $\xi$ becomes
\begin{equation}
\xi=R^{-1/r}f_{Rt}(tR^{z/r}[1+g^{\nu}\ln R/(ar)],D^{-1} R^{-1/r\kappa}),
\label{eq:ftst}
\end{equation}
upon replacing Eq.~(\ref{eq2}) for $g$ with Eq.~(\ref{tb}) for $t$ and setting again $b=R^{-1/r}$, where $f_{Rt}$ is a scaling function. Using Eq.~(\ref{brz1}), Eq.~(\ref{eq:ftst}) can also be written as
\begin{equation}
\xi=R^{-1/r}f_{Rt}(tR^{z/r}R^{\nu(r-z)/[r(1+\nu)]},D^{-1} R^{-1/r\kappa}).
\label{eq:ftstr}
\end{equation}
Equation~(\ref{eq:ftst}) gives rise to a new driven time scale
\begin{equation}
\tau_R=R^{-z/r}\left(1+\frac{g^\nu}{ar}\ln R\right)^{-1},\label{tr}
\end{equation}
instead of the usual $R^{-z/r}$, which is obtained from a similar procedure except for replacing Eq.~(\ref{tb}) with $t(b)=tb^{-z}$. From Eq.~(\ref{eq:ftstr}), the new driven time scale is also given by
\begin{equation}
\tau_R=R^{-z/r}R^{\nu(z-r)/[r(1+\nu)]},\label{trrr}
\end{equation}
viz., there is an additional $R$ factor. Using Eqs.~(\ref{eq1}) and~(\ref{eq:ftst}), one can rewrite Eq.~(\ref{tr}) into
\begin{equation}
\tau_R=R^{-z/r}\ln(\xi/\xi_0)/\ln(f_{Rt}/\xi_0),\label{trx}
\end{equation}
which contains an additional $\ln(\xi/\xi_0)\propto g^{-\nu}$ and thus resembles somehow Eq.~(\ref{teq}) for $\tau_{\xi}$ if the first argument of $f_{Rt}$ is vanishingly small and $D$ sufficiently large.

Finally, we study the KZ scaling for $D=\infty$. On the one hand, for a usual continuous phase transition with a simple critical point, the first argument of $\xi$ on the righthand side of Eq.~(\ref{eq3}) is replaced by Eq.~(\ref{gbus}) and that of $f_R$ in Eq.~(\ref{eq:fts}) is replaced by $gR^{-1/r\nu}$ with $r$ given by Eq.~(\ref{rznu})~\cite{Zhong1,Zhong2,Huang}. Accordingly, $\xi\propto R^{-1/r}$ for $gR^{-1/r\nu}={\rm constant}$ in general and $\hat{g}R^{-1/r\nu}=\pm1$ in particular. The latter defines the frozen $\hat{g}=\pm R^{1/r\nu}$ or the frozen instant $\hat{t}=\hat{g}/R=\pm R^{-z/r}$ using Eq.~(\ref{rznu}). The first condition leads to $R^{-1/r}=\hat{\xi}\sim\hat{g}^{-\nu}$ and $R^{-z/r}=\hat{\tau}\sim\hat{g}^{-\nu z}$, which dictate that the frozen condition is the driven length scale $R^{-1/r}$ and the driven time scale $R^{-z/r}$ equal the correlation length and correlation time, respectively, ignoring the proportional constants. The second condition then requires the ``time to the critical point"~\cite{KZ2,KZ3,Dziarmaga,Polkovnikov,inexper4} match the other two time scales. Therefore, the density of the topological defects formed at $\hat{t}$ after the transition is proportional to $\hat{\xi}^{-d}\sim R^{d/r}$ in a $d$-dimensional space.

For the KT phase transition on the other hand, it is now at
\begin{equation}
\hat{g}\left(1+\frac{\hat{g}^{\nu}}{ar}\ln R\right)^{-1/\nu}=C,\label{frozen}
\end{equation}
that $\xi= R^{-1/r}f_R(C,0)$, which is exact in the sense that there exists no other dependence on $g$ and $R$ through the scaling function $f_R$ provided that other sub-leading and corrections to scaling which are not considered are neglected, where $C$ is a constant. The reason that we choose a constant $C$ instead of simply 1 is that $\xi$ contains now a dimensional proportional constant $\xi_0$. Equation~(\ref{frozen}) gives rise to
\begin{eqnarray}
R^{-1/r}=\hat{\xi}/\xi(C)=\xi_0\exp(a\hat{g}^{-\nu})/\xi(C),\qquad\qquad\label{trxi}\\
R^{-z/r}=\hat{\xi}^z/\xi(C)^z=\tau(g)/\tau(C)=\tau_0\exp(za\hat{g}^{-\nu})/\tau(C),\nonumber\\
\label{trte}
\end{eqnarray}
where $\tau_0=\xi_0^z$ is a microscopic time scale and $\tau(C)=\xi(C)^z$ is the correlation time at $g=C$, a constant, as used in Sec.~\ref{stog}, by turning the asymptotic relation in Eq.~(\ref{txiz}) into an equation. In other words, the correlation length and the correlation time are now proportional to their corresponding driven ones. Note, however, that $r$ is given by Eq.~(\ref{eq6}) and hence depends on $g$ and $R$. Combining Eqs.~(\ref{frozen}) and~(\ref{eq6}) and using Eqs.~(\ref{grt}) and~(\ref{trte}), one finds
\begin{equation}
\hat{t}=\hat{g}/R=CR^{-z/r}=C\tau_0\exp(za\hat{g}^{-\nu})/\tau(C),\label{that}
\end{equation}
which agrees with previous results up to a proportional constant~\cite{Dzia,Gardas,Deutsch}. Therefore, the KZ scaling still holds in that the defect density is again proportional to $R^{d/r}$, though with an exponent $r$ depending on $R$ rather than a constant. This is the observation of previous studies~\cite{Braun,Dzia,Gardas,Weiss,Jelic,Deutsch}. However, we note that this must be observed at the particular $\hat{g}$ for the KZ mechanism with its corresponding $r$ solved from Eq.~(\ref{frozen}) or~(\ref{that}) and Eq.~(\ref{eq6}) with $\hat{g}$ in place of $g$ for a given $R$. In fact, Eq.~(\ref{that}) is solved by
\begin{equation}
\hat{g}=\left[\frac{a z\nu}{W(-az\nu[C\tau_0R/\tau(C)]^{-\nu})}\right]^{1/\nu},\label{gw}
\end{equation}
where $W$ is the Lambert W function. Plugging this into Eq.~(\ref{eq6}) then leads to $r$. For measurements at arbitrary $g$, $r$ depends on it as well as $R$ and is given directly by Eq.~(\ref{eq6}). Moreover, as indicated, the scaling function $f_R$ also brings additional dependence on $g$ and $R$.

At the frozen condition Eq.~(\ref{frozen}) we chosen, the three times, Eqs.~(\ref{trte}) and~(\ref{that}), are proportional. one can of course choose $C/\tau(C)=1$, or $C=[az\nu/W(az\nu\tau_0^{-\nu})]^{1/\nu}$, in Eq.~(\ref{that}) and the resultant equation is just the frozen condition employed previously~\cite{Dzia,Gardas,Deutsch}. At this condition, Eqs.~(\ref{trxi}),~(\ref{trte}), and~(\ref{that}) become $\hat{t}=\tau(\hat{g})=\tau_0\exp(za\hat{g}^{-\nu})=CR^{-z/r}$ and $C^{1/z}R^{-1/r}=\hat{\xi}$.

For the case of the anomalous relaxation time in the form Eq.~(\ref{teq}), because the FTS form is still Eq.~(\ref{eq:fts}), the frozen condition is still given by Eq.~(\ref{frozen}), which again leads to Eq.~(\ref{trxi}) for the proportionality of the driven length and correlation length and Eq.~(\ref{trte}), though $r$ is given by Eq.~(\ref{brz1}) instead of Eq.~(\ref{eq6}). However, the latter, Eq.~(\ref{trte}), does not now represent the equality of the driven time and the relaxation time according to Eqs.~(\ref{tr}) and~(\ref{teq}). Yet, one may multiply Eq.~(\ref{trte}) directly by $\ln(\xi/\xi_0)$. The left-hand side then becomes $\tau_{\xi}$ whereas the right-hand side is $\tau_R$, Eq.~(\ref{trx}), up to a constant to the leading order in which $f_{Rt}\simeq f_{Rt}(0,0)$, i.e., the relaxation time is approximately equal to the driven time. In fact, at the frozen condition, Eq.~(\ref{frozen}), using Eqs.~(\ref{tr}),~(\ref{trte}),~(\ref{teq}), and~(\ref{eq1}), one finds
\begin{equation}
\tau_R=\tau_0C^{\nu}\hat{g}^{-\nu}\exp\left(z a \hat{g}^{-\nu}\right)/\tau(C)\simeq C^{\nu}\tau_{\xi}/[a\tau(C)].\label{trte1}
\end{equation}
Moreover, Eqs.~(\ref{frozen}) and~(\ref{brz1}) together with Eqs.~(\ref{grt}) and~(\ref{tr}) lead to
\begin{equation}
\hat{t}=CR^{-z/r}\left(1+\frac{\hat{g}^\nu}{ar}\ln R\right)^{-1}=C\tau_R.\label{that1}
\end{equation}
Therefore, at the frozen condition, the time to the critical point, the driven time and the relaxation time are all exactly proportional. That $\hat{t}\propto\tau_{\xi}$ agrees up to the proportional constant with the condition employed in Ref.~\cite{Jelic}. In addition, $\hat{g}$ is again solved from Eqs.~(\ref{trte1}),~(\ref{that1}), and~(\ref{grt}) by
\begin{equation}
\hat{g}=\left[\frac{a z\nu}{(1+\nu)W(-az\nu (C_2\tau_0R)^{-\nu/(1+\nu)}/(1+\nu))}\right]^{1/\nu},\label{gw1}
\end{equation}
in place of Eq.~(\ref{gw}), and $r$ is given by Eq.~(\ref{brz1}) in place of Eq.~(\ref{eq6}) with $g$ replacing by $\hat{g}$, where $C_2=C^{\nu+1}/[a\tau(C)]$. Of course, selecting $C_2=1$ or $C=[az\nu/(1+\nu)W(az\nu\tau_0^{-\nu/(1+\nu)}/(1+\nu))]^{1/\nu}$ one then has $\hat{t}=\tau_{\xi}=C\tau_R$ in place of Eqs.~(\ref{trte1}) and~(\ref{that1}).

We finally remark that the present scaling theories contain a non-universal constant $a$ in an essential way besides the usual universal critical exponents as the usual scaling theories do. This is in contrast with the usual non-universal multiplying constants for each scaled variable in scaling functions since they are negligible. However, this constant can be determined, for example, from data collapses.

\section{Model and method\label{model}}
The Hamiltonian of 1D BH model is~\cite{fisher}
\begin{equation}
\widehat{{\cal H}}=-J\sum_{j}(\widehat{a}^{\dagger}_j\widehat{a}_{j+1}+H.c.)-\mu\sum_{j}\widehat{n}_j+\frac{U}{2}\sum_{j}\widehat{n}_j(\widehat{n}_j-1)
\label{HM}.
\end{equation}
where \(\widehat{a}^{\dagger}_j\) (\(\widehat{a}_j\)) creates (annihilates) a boson at site $j$, \(\widehat{n}_{j}=\widehat{a}^{\dagger}_{j}\widehat{a}_{j}\) is a number operator at site $j$, $J$ represents the hopping amplitude, \(\mu\) stands for the chemical potential, and $U$ denotes an on-site interaction. We set $U=1$ and the maximum occupation number \(n_{\rm max}=3\), which is sufficient for the calculation at unit filling here. For a large $J$, the bosons can hop freely in the lattice and the system falls in a superfluid phase; whereas for a small $J$, hopping is energetically costly and a Mott insulator phase results. The KT phase transition between the two phases appears at the tip of Mott insulator lobes~\cite{fisher,Pino,Krutitsky}. In particular, for unit filling, the critical point falls on the line
\begin{equation}
\mu(J)=-1.538J+0.559\label{muj}
\end{equation}
to a very good approximation~\cite{Pino}. Accordingly, to approach the critical point, both $\mu$ and $J$ have to be varied coordinately along the line. The distance to the critical point should also be measured along the line. However, since $\mu$ and $J$ are related, we can nevertheless define
\begin{equation}
g=J_c-J,\label{gjkt}
\end{equation}
for the critical point at $J_c$, since $\mu$ contributes through the line a proportional constant to $g$ and hence can be absorbed in the constant $a$ in Eq.~(\ref{eq1}). In the linear driving, we increase $J$ as $J(t)=J_c+Rt$ for $R>0$ from the Mott phase with $J<J_c$ at $t<0$ to the superfluid phase. In addition, the dynamic critical exponent $z=1$ both in the superfluid phase and the KT transition~\cite{fisher}. The central charge $c=1$ and hence $\kappa\simeq1.3441$ from Eq.~(\ref{kappa}). Also $\nu=1/2$ as mentioned.

To study the KT phase transition in the model~(\ref{HM}), we employ the iTEBD algorithm to solve it~\cite{Vidal,Vidal03,Vidal04,Vidalitebd}.
Its basic idea is to cast the wave function of the system into a matrix product state~\cite{Klumper,Verstraete1,Schollwock} through Vidal decomposition~\cite{Vidal03}, viz.,
\begin{equation}
|\Psi\rangle=\sum_{i_1\cdots i_N \atop\alpha_1 \cdots \alpha_{N-1}} \Gamma_{\alpha_1}^{[1]i_{1}} \lambda_{\alpha_1}^{[1]} \Gamma_{\alpha_1\alpha_2}^{[2]i_{2}} \lambda_{\alpha_2}^{[2]} \cdots \Gamma_{\alpha_{N-1}}^{[N]i_{N}} \left|i_{1}\cdots i_{N}\right\rangle,\label{psi}
\end{equation}
for a lattice of $N$ sites, each with a local basis $|i_j\rangle$ ($j=1,\cdots,N$) of ${\cal N}$ degrees of freedom, with open boundary conditions. Here $\lambda_{\alpha_j}^{[j]}$ and $\Gamma_{\alpha_{j-1} \alpha_j}^{[j]i_{j}}$ are local vectors and tensors on site $j$ and represent the Schmidt coefficients and the relation between Schmidt states and the local bases, respectively, in a series of Schmidt decomposition. In addition, the bond indices $\alpha_j$ run from 1 to the bond dimension $D$. The evolution of the wave function is achieved by updating the local $\Gamma$ and $\lambda$ according to an evolution operator $\widehat{{\cal U}}$ for the local Hamiltonian $\widehat{{\cal H}}$. This operator can be either $\widehat{{\cal U}}(t)=\exp(-\widehat{{\cal H}}t)$ for an imaginary-time evolution or $\widehat{{\cal U}}(t) = \exp(-i\widehat{{\cal H}}t)$ for a real-time evolution. In practice, we expand $\widehat{{\cal U}}(t)$ via the Suzuki-Trotter method to the second order and choose the time interval to be $\delta t=0.01$. This approximation is the main origin of errors~\cite{Vidal04,Hatano,Garcia}. For an infinite lattice with translational invariance of a cell of two sites denoted as $A$ and $B$, $\Gamma^{[2j]}=\Gamma^{A}$, $\lambda^{[2j]}=\lambda^{A}$, $\Gamma^{[2j+1]}=\Gamma^{B}$, and $\lambda^{[2j+1]}=\lambda^{B}$ for all $j$. Consequently, only the states on sites A and B need to be updated~\cite{Vidalitebd}.

From Eq.~(\ref{psi}), the entanglement entropy between site $A$ and site $B$ is determined from the von Neumann entropy of either site $A$ or site $B$, denoted as $A/B$, as
\begin{equation}
	S = -\sum_{\alpha=1}^{D}\left(\lambda_{\alpha}^{A/B}\right)^{2}\ln \left(\lambda_{\alpha}^{A/B}\right)^{2}.\label{ee}
\end{equation}
In principle, $S$ computed from $A$ and $B$ is identical. However, their numerical results may be slightly different and we average them. On the other hand, to obtain the correlation length $\xi$, one can define a left transfer matrix~\cite{CLcal}
\begin{equation}
	{\cal T}_{\alpha \alpha';\beta \beta'} = \sum_{i}^{{\cal N}}\lambda_{\alpha}^{A/B} \lambda_{\alpha'}^{A/B}\Gamma_{\alpha \beta}^{A/B i}\left(\Gamma_{\alpha' \beta'}^{A/B i}\right)^{*},
\end{equation}
and a corresponding right one, where the star denotes complex conjugation. For $\Gamma$ and $\lambda$ in a canonical form, the largest eigenvalue of ${\cal T}_{\alpha \alpha';\beta \beta'}$ is unity~\cite{Vidalitebd,CLcal}. Accordingly, the second largest (in terms of absolute value) eigenvalue $\epsilon_{2}$ determines the largest correlation length
\begin{equation}
	\xi=-\frac{1}{\log \left|\epsilon_{2}\right|}.\label{xie}
\end{equation}
One sees therefore that $S$ and $\xi$ are computed from different sources though they are related by Eq.~(\ref{scx}).

In the imaginary-time evolution, we start with an initial ordered state which is a direct product of local states that are, in turn, superposition of all possible occupation states with equal probability. This highly ordered state ought to reduce most, if not eliminate, initial distribution of vortices and qualifies our neglect of the correlation time $\tau_{\xi}$. Indeed, we have found that the logarithmic correction in $\tau_{\xi}$ appears unnecessary. We then solve the model with the iTEBD method at a given $J$ and hence $\mu$ determined by Eq.~(\ref{muj}). In the real-time driving, the initial state is prepared by the same procedure until its equilibration and then driven at a given $R$ through $J_c$ to the superfluid phase.

\section{Numerical results\label{num}}
\subsection{\label{stcd}Short-imaginary-time critical dynamics and the KT point}
To verify Eq.~(\ref{eq:four}) for the short-time critical dynamics, we employ the imaginary-time evolution of $\xi$ and $S$. The first thing to do is to fix $J_{c}$. This has been studied extensively~\cite{Krutitsky} and a recent one employed a finite entanglements~\cite{Pino}. Here we apply the short-time critical dynamics embodied in Eq.~(\ref{eq:four}) and, in particular, Eq.~(\ref{eq4}), to estimate $J_c$.

\begin{figure}
\includegraphics[width=1\columnwidth]{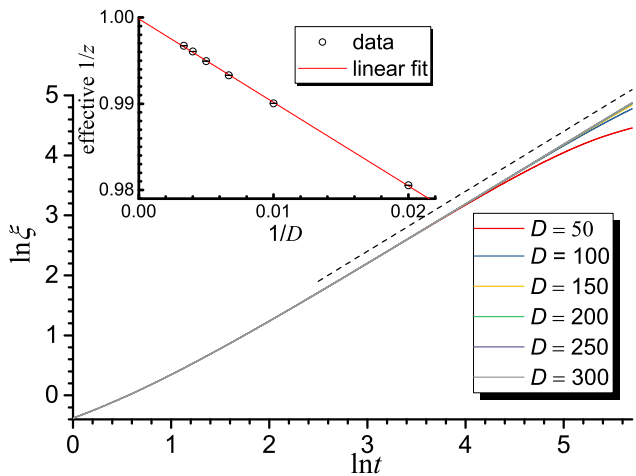}
\caption{(Color online) Imaginary-time evolution of the correlation length $\xi$ for different bond dimensions \(D\) given in the legend at \(J=0.3\). The slope of the dashed line is exactly $1/z=1$. The inset shows the dependence of the effective \(1/z\) on \(1/D\) and its extrapolation. The errorbars are far smaller than the symbol sizes.}
\label{fig1}
\end{figure}
From Eq.~(\ref{eq4}), exactly at $J=J_c$ and for $D=\infty$, the slope of $\ln\xi$ versus $\ln t$ must be $1/z=1$, because $z=1$ strictly for the BH model~\cite{fisher} as mentioned above. This is the method employed previously in the short-time critical dynamics to determine the critical point~\cite{ZhengXY,Ying,Zheng,Lei}. However, for a finite $D$, the scaling function $f_t(0, D^{-1}t^{1/z\kappa})$ matters and the effective slope is not 1. This is also true if $g\neq0$ and the first argument of $f_t$ also play a role. Our strategy is to find the smallest $J$ for which $z=1$ after extrapolated to an infinite $D$ since $z=1$ in the superfluid phase. This is illustrated in Fig.~\ref{fig1} for a specific $J=0.3$ that were assumed tentatively to be the critical point.

\begin{figure}
\includegraphics[width=1\columnwidth]{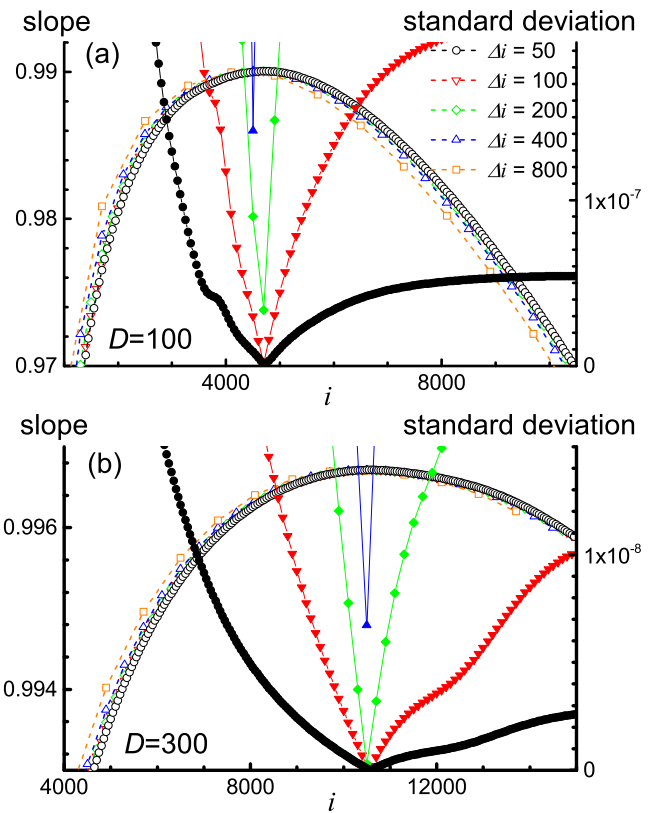}
\caption{(Color online) \label{sst} Slopes (left axis) and standard deviations (right axis) of least-squared linear fits to $\ln\xi$ versus $\ln t$ for five step sizes $\Delta i$ given in the legend (shared by two panels) at (a) $D=100$ and (b) $D=300$ at $J=0.3$. $i$ is the number of steps and is related to the imaginary time by $t=i\times\delta t$ with $\delta t=0.01$. Opened symbols connected by dashed lines denote slopes (which exhibit maxima) and filled symbols connected by solid lines stand for standard deviations (which show valleys). Lines connecting symbols are only a guide to the eyes.}
\end{figure}
In Fig.~\ref{fig1}, one sees that the correlation length $\xi$ is almost a straight line with a slope close to 1 for sufficiently long time. When $\xi$ grows to the imaginary time at which it is about the size of $D^{\kappa}$, whose logarithm is about $5.2$, $6.2$, $6.7$, $7.1$, $7.4$, and $7.7$ for the $D$ values listed in Fig.~\ref{fig1}, the system enters the finite-entanglement scaling regime governed by Eq.~(\ref{fes}) and the line curves down, since it is then $D$ rather than $t$ that is the controlling factor. Although the lines of different $D$ values look coincident for $t$ not too late, their slopes are slightly different. In addition, they are not strictly straight lines; different ranges of time exhibit slightly different slopes. In Fig.~\ref{sst}, we show the slopes and the standard deviations of least-squared linear fits to the $\ln\xi$ versus $\ln t$ curves for $D=100$ and $300$. One sees that the slopes of different step sizes $\Delta i$ peak at almost the same instant, with a variation of no more than several step sizes, though the peaks are broad and delayed for the large $D$. Moreover, the standard deviations of the fits fall to their valley bottoms concurrently, with possibly a similar variation. This indicates that the maximum slope corresponds to the straightest part of the curve. Accordingly, we identify the maximum slopes as the effective $1/z$. Generally, the standard deviations decrease with the step sizes used in the fits. Yet, for some $D$ and $J$, those for $\Delta i=50$ and $100$ may invert. To avoid too small a step size, we thus present the results for $\Delta i=100$. The results for these two step sizes are, in fact, almost identical, though, those for bigger step sizes vary slightly as the standard deviations rise. The effective $1/z$ so obtained for different $D$ is drawn in the inset of Fig.~\ref{fig1}. Their relation appears well linear, reminiscent somehow of a first-order expansion of $f_t$ in $D^{-1}t^{1/z\kappa}$. The intercept at $D=\infty$ is $1/z$, which is \(0.99983(8)\) for the given $J=0.3$ under the assumption that it be $J_c$. The same result is found for $\Delta i=50$, while it is \(0.99985(7)\) for $\Delta i=200$, showing the small variation with $\Delta i$ for these step sizes. The tiny difference from 1 of the extrapolated $1/z$ indicates that $0.3$ is not exactly $J_c$: The first argument of $f_t$ is not exactly zero and still matters. We note that the standard errors of the effective $1/z$ are extremely small because of the small fitted ranges of time. Also the linearity is rather good. Consequently, the standard error of the extrapolated $1/z$ is small, down to the fifth decimal place, even though the critical line, Eq.~(\ref{muj}), has only three significant digits and usually the same number of digits is kept for the second-order Suzuki-Trotter expansion~\cite{Cao}.

\begin{figure}
	\includegraphics[width=1\columnwidth]{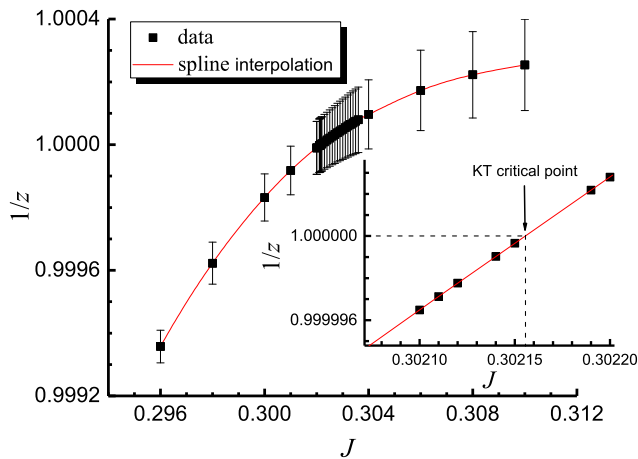}
	\caption{\label{fit}(Color online) The extrapolated \(1/z\) for various \(J\) in the short-time critical dynamic regime. The inset magnifies the part near $1/z=1$.}
\end{figure}
Now, we compute $\xi$ for various $J$ and $D$ and repeat the above procedure for each $J$. The resultant extrapolated $1/z$ is shown in Fig.~\ref{fit}. Its relation with $J$ is definite, even though the errorbars are now visible due to the small scale and lengthen moderately as $J$ increases. From the inset of Fig.~\ref{fit} alone, the KT critical point is \(J_c=0.302156(5)\). For the step size $\Delta i=50$ and $200$, we find \(J_c=0.302228(5)\) and $0.301884(5)$, respectively. Upon taking into account possible errors in the critical line, Eq.~(\ref{muj}), and the Suzuki-Trotter expansion, our conservative estimate is \(J_c=0.302(1)\). In fact, the $1/z$ of the lower bound just touches 1 with its errorbar in Fig.~\ref{fit}. This $J_c$ agrees with many previous estimates~\cite{Pino,Krutitsky} and especially two recent ones~\cite{Pino,Arcila}. We note that \(1/z<1\) for \(J <J_c\) does not mean $z>1$ in the Mott phase. Rather, as mentioned above, it implies that the first argument of $f_t$ is not zero and its contribution diminishes as the critical point is approached. Only exactly at \(J =J_c\) can the related argument be ignored and the extrapolated $z=1$. Similarly, \(1/z>1\) for \(J >J_c\) does not mean $z<1$ in the superfluid phase; its $z$ equals strictly 1. In fact, the errorbars cover $1$ for not too large $J$ and the rise of the extrapolated $1/z$ with $J$ gets somehow sluggish for \(J >J_c\). However, in the gapless superfluid phase, every $J$ is a critical point. Accordingly, the first argument must be zero for each $J>J_c$, even though the second argument may well be given by the same form as the one in $f_t$, irrespective of its $\kappa$ value. This seems to indicate that there exists another relevant variable which is not considered in the present theory for the superfluid phase. We leave this for future study.

\begin{figure*}
	\includegraphics[width=\textwidth]{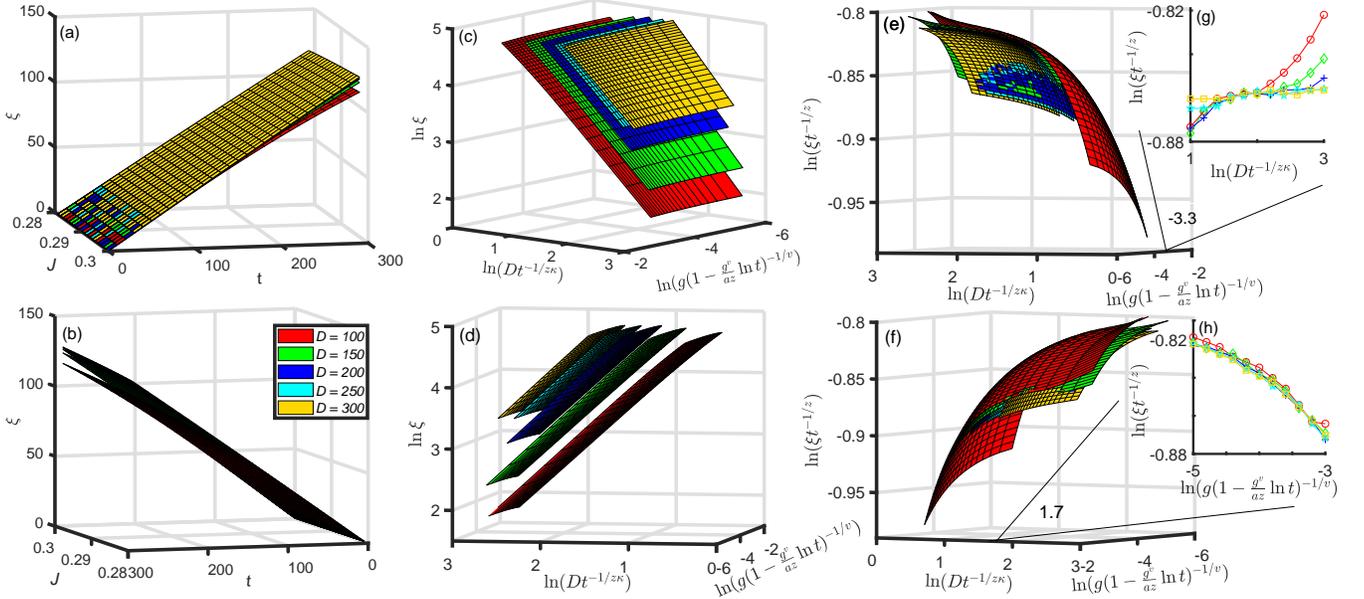}
\caption{\label{fig:epsart3} (Color online) (a) and (b) Imaginary-time evolution of the correlation length $\xi$ for various $J$ and $D$. (c) and (d) Unscaled, and (e) and (f) scaled correlation length according to Eq.(\ref{eq4}). The two rows of panels are different only in the angle of view. The two insets, (g) and (h), display sections of the main plots at $\ln\{g[1-g^\nu\ln t/(az)]^{-1/\nu}\}=-3.3$ and $\ln(Dt^{-1/z\kappa})=1.7$, respectively. Symbols are 2D interpolations and lines connecting them are only a guide to the eyes. All panels share the same legend in (b).}
\end{figure*}
Having fixed the critical point, we can corroborate it through scaling collapses and verify accordingly the theory. We have to collapse three-dimensional surfaces because we have to consider the bond dimension $D$ besides others. The imaginary-time evolution of the correlation length $\xi$ for various $J$ and $D$ is shown in Fig.~\ref{fig:epsart3}(a) and~\ref{fig:epsart3}(b) for two different angles of view. The separation of $\xi$ for different $D$ at late time and the slight rise of its slopes with increasing $J$ below $J_c$ are visible. In terms of the scaled variables
\begin{equation}
X=\ln\left[g\left(1-\frac{g^\nu}{az}\ln t\right)^{-1/\nu}\right] {\rm and~} Y=\ln\left(Dt^{-1/z\kappa}\right),\label{xy}
\end{equation}
$\ln\xi$ appears as parallel surfaces for different $D$ within the ranges of the coordinates as presented in Fig.~\ref{fig:epsart3}(c) and~\ref{fig:epsart3}(d). Yet, after $(\ln t)/z$ in Eq.~(\ref{eq4}) is deducted, the surfaces collapse onto each other for $-3.5\lesssim X \lesssim-2$ and \(1.3\lesssim Y\lesssim2.5\) as the color mixing in Figs.~\ref{fig:epsart3}(e) and~\ref{fig:epsart3}(f) manifest. This can also be seen from the two insets, Figs.~\ref{fig:epsart3}(g) and~\ref{fig:epsart3}(h), where curves of different $D$ collapse within some ranges. We note however that the sections as demonstrated in Figs.~\ref{fig:epsart3}(g) and~\ref{fig:epsart3}(h) are only approximate representations of the 3D plots, since the symbols have to be obtained from 2D interpolations which contains errors. In addition, the surfaces in the region for smaller $X$ within $-6\lesssim X \lesssim-3.5$ for $1.5\lesssim Y \lesssim2$ are also very close to each other.
Note that $Y$ is the inverse of the corresponding scaled variable in $f_t$ and thus the overlap region is indeed the short-time regime in which $\exp(X)\ll 1$ and $\exp(-Y)\ll 1$. For smaller $Y$ values, as mentioned above, the system enters the finite-entanglement scaling regime and the surface bends down, as the surface of $D=100$ exhibits. However, we note that both the short-time regime and the finite-entanglement regime can be described by the same scaling form even though their leading behavior is different~\cite{Huang,Cao}. One reason that the overlap region is limited stem from the fact that the scaling theory is built on Eq.~(\ref{eq1}). Corrections as given by Eq.~(\ref{eq1c}) are needed far away from the critical point~\cite{Amit,Balog,Pelissetto}. They will appear as extra terms to $X$. In addition, other corrections to scaling arising for example from small $D$ have not been taken into account in the present theory~\cite{Wegner}. Nevertheless, the good collapses show that the exponential form of the correlation length, Eq.~(\ref{eq1}), is valid at least for some regions of the parameter spaces and confirm the resultant short-time dynamic scaling in imaginary time of the KT phase transition.

In the above surface collapses, we need to know the parameter $a$ as exemplified in Eq.~(\ref{xy}). It can be obtained by adjusting it to achieve the best collapses. Yet, we find that the quality of the collapses does not depend sensitively on a small adjustment of $a$, possibly due to the difficulty in assessing three-dimensional collapses. Our result is $a=2$, in agreement with previous results~\cite{Gardas}. When higher order terms such as $a_1$ in Eq.~(\ref{eq1c}) are taken into account, one has to adjust in addition those parameters along with $a$. We leave this for future study.

\begin{figure}[b]
	\includegraphics[width=1\columnwidth]{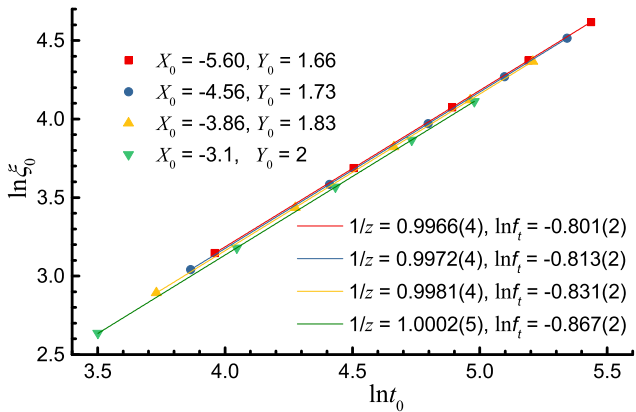}
	\caption{\label{fig:epsart2} (Color online) The correlation length $\xi_0$ and imaginary time $t_0$ at four given pairs of $(X_0, Y_0)$. Lines are linear fits to the data of each pair.}
\end{figure}
We now examine further the theory as the 3D surface collapses are not so intuitive. In our estimate of $J_c$, we have to deliberately nullify the effects of the two scaled variable $X$ and $Y$, Eq.~(\ref{xy}), separately in order to acquire the unit slope of $1/z$. In fact, this goal can also be achieved for constant $X$ and $Y$ so that $f_t(\exp(X),\exp(-Y))$ is also a constant.
Thus, given a $Y=Y_0$ one has, for each $D$, a $t_0$ (not to be confused with the same symbol used in Sec.~\ref{tstcd}), which results in a $g$ for a given $X=X_0$ using Eq.~(\ref{xy}). The pair of given $(X_0,Y_0)$ leads to an $\ln\xi_0$ from Fig.~\ref{fig:epsart3}(c) and~\ref{fig:epsart3}(d). All these $t_0$ and $\ln\xi_0$ for a given $(X_0,Y_0)$ pair must then obey Eq.~(\ref{eq4}), viz., $\ln\xi_0$ versus $\ln t_0$ must be a straight line with a slope of $1/z=1$ and an incept of $\ln f_t(\exp(X_0),\exp(-Y_0))$. In Fig.~\ref{fig:epsart2}, we depict the results for four pairs of $(X_0,Y_0)$.
One sees from Fig.~\ref{fig:epsart2} that the slopes are close to but not 1 even taking the standard errors into account for pairs of points outside the color-mixed region while they are 1 in the region. There are two main uncertainties for the results. One arises from possible small corrections and the other from the errors in the 2D interpolation of $\ln\xi_0$ for a given $(X_0,Y_0)$. Note that the four fitted lines do not overlap onto each other; they have different incepts, which are consistent with the corresponding values of $\ln(\xi t^{-1/z})$ in Fig.~\ref{fig:epsart3}(e) and~\ref{fig:epsart3}(f).

\begin{figure}[b]
	\includegraphics[width=1\columnwidth]{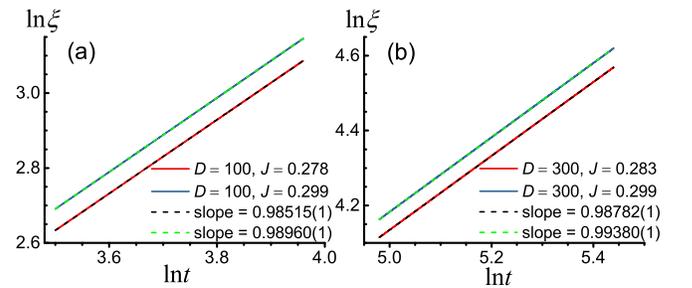}
	\caption{\label{fig:XYJ} (Color online) $\ln\xi$ versus $\ln t$ within the imaginary-time ranges covered by the lines for two $J$ values listed and (a) $D=100$ and (b) $D=300$. Solid lines are numerical solutions of the BH model while dashed lines are linear fits.}
\end{figure}
\begin{figure*}
	\includegraphics[width=\textwidth]{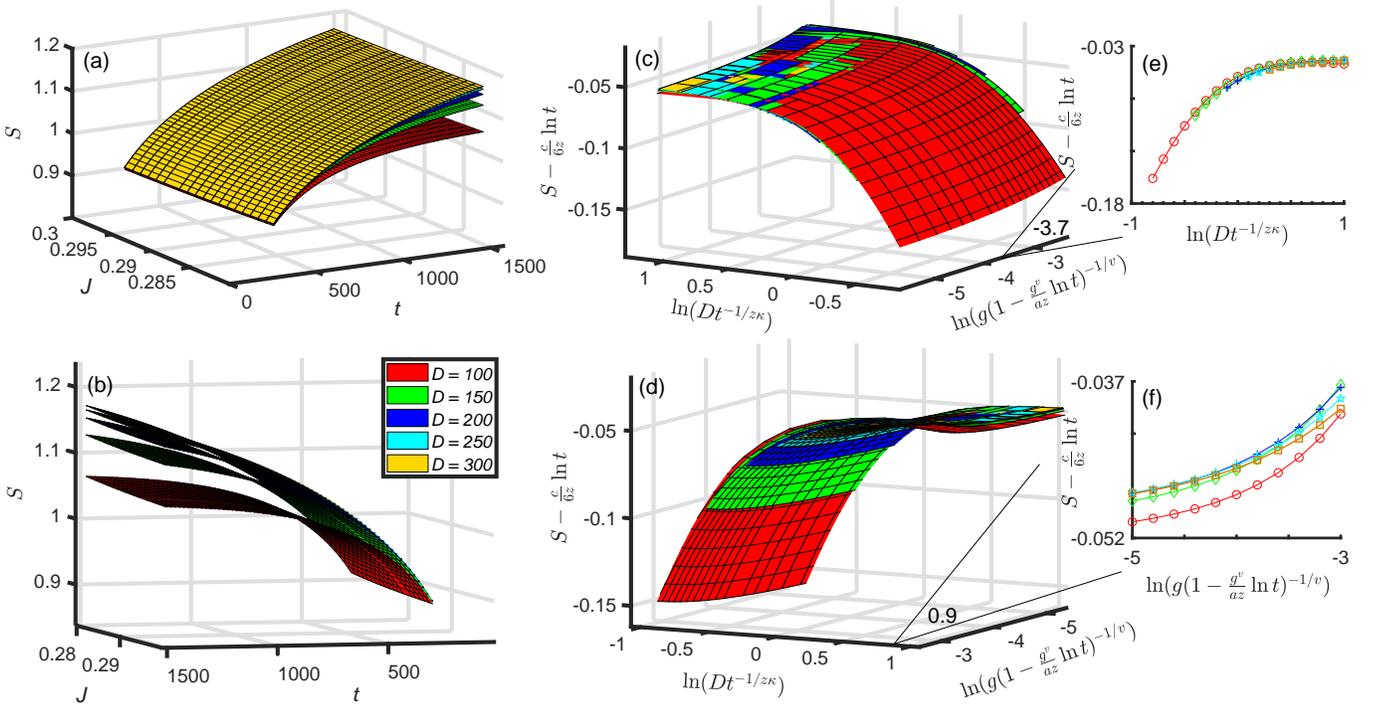}
	\caption{\label{fig:entim} (Color online) (a) and (b) Imaginary-time evolution of the entanglement entropy $S$ for various $J$ and $D$. (c) and (d) Short-time dynamic scaling of $S$. The two rows of panels are different only in the angle of view. The two insets, (e) and (f), display sections of the main plots at $\ln\{g[1-g^\nu\ln t/(az)]^{-1/\nu}\}=-3.7$ and $\ln(Dt^{-1/z\kappa})=0.9$, respectively. Symbols are 2D interpolations and lines connecting them are only a guide to the eyes. All panels share the same legend in (b).}
\end{figure*}
We emphasize that even though the slopes far away from the apparent color-mixed region are not 1, these values represent a significant improvement to those direct fits without consideration of the scaled variables in such a relatively wide range. One might think that the fitted slopes for the given pairs were not good enough, as, for example, compared with those in the inset of Fig.~\ref{fig1} and, especially, those in Fig.~\ref{fit}. However, they are obtained from completely different conditions. The $1/z$ in Fig.~\ref{fit} was extrapolated to infinite $D$, while the best one in Fig.~\ref{fig1} is about $0.9967$ for the straightest and largest from a small range of time and for a $J$ close to $J_c$ and the largest $D=300$. For a comparison, we display in Fig.~\ref{fig:XYJ} the fitted slope of the effective $1/z$ for the maxima and minima $g$ and $D$ and for the same ranges of time as those drawn in Fig.~\ref{fig:epsart2}. The parameters are reckoned as follows. For the minimum $D=100$ used in Figs.~\ref{fig:epsart3} and~\ref{fig:epsart2}, the range of $\ln t$ is $3.50$--$3.96$ for the four selected $Y_0$ values. Meanwhile, the four selected $X_0$ values together with the above $\ln t$ range give rise to a $J$ range of $0.278$--$0.299$. Similarly, for the maximum $D=300$, one finds the ranges of $\ln t$ and $g$ are $4.98$--$5.44$ and $0.283$--$0.299$, respectively. From Fig.~\ref{fig:XYJ}, one sees that the fitted effective $1/z$ of the one closest to 1 for the biggest $g$ and larger $D=300$ is smaller than $0.994$, still inferior to the smallest one in Fig.~\ref{fig:epsart2}. Moreover, note that each slope in Fig.~\ref{fig:XYJ} only contains one $D$, whereas those of Fig.~\ref{fig:epsart2} also involve others. Therefore, the theory does take the effects of the distance to the critical point and the bond dimension into account and does improve the results.

To further substantiate the scaling theory, we present the imaginary-time evolution of the entanglement entropy $S$ and its scaling in Fig.~\ref{fig:entim}, again in two different angles of view. From Figs.~\ref{fig:entim}(a) and~\ref{fig:entim}(b), it is seen that $S$ increases with the imaginary time for a given $J$, while very slightly decreases as $J$ increases. The former is due to the growth of the correlation length with time and the latter is opposite to the trend of $\xi$ exhibited in Fig.~\ref{fig:epsart3}(a). This is a consequence of the competition between the increase of $\xi$ with $g$ and the equilibration time. For $\xi$, it increases with $J$ as the former wins, whereas for $S$, the logarithm of $\xi$ to which it is proportional renders it faintly decreasing. Also, the bigger the bond dimension $D$, the more the entanglement is kept and hence $S$ rises. Different $S$ surfaces separate significantly for $t>500$. However, after being scaled by Eq.~(\ref{eq:iment}), such widely separated surfaces in Figs.~\ref{fig:entim}(a) and~\ref{fig:entim}(b) become significantly close and almost combine to one surface as shown in Figs.~\ref{fig:entim}(c) and~\ref{fig:entim}(d). The two orthogonal sections in Figs.~\ref{fig:entim}(e) and~\ref{fig:entim}(f) also show quite good collapses. Only the red curve for $D=100$ appears somehow poor due to the small scale and its relatively small $D$. As pointed out in Sec.~\ref{model}, $S$ is computed from a different source to $\xi$ though they are related by Eq.~(\ref{scx}). Therefore, the nearly collapse of the $S$ surfaces supports the scaling theory.

However, one sees from Figs.~\ref{fig:entim}(c) and~\ref{fig:entim}(d) that there is a small gap between surfaces of different $D$ values. They even cross over each other near $\ln (D t^{-1/z\kappa})=0.5$ as can be seen from the order of the colors from up to down on both sides. The reason is believed to result from accumulative errors from normalization. The iTEBD algorithm we employed for the imaginary-time evolution is not normalized every step~\cite{Schollwock}. The resultant errors for one eigenvalue of the transfer matrix in computing the correlation length $\xi$ from Eq.~(\ref{xie}) may be small. However, the computation of the entanglement entropy, Eq.~(\ref{ee}), involves all the Schmidt coefficients. As a consequence, the accumulative errors may be large and lead to the small gap in the scaled surfaces.

\subsection{\label{ftswd}FTS with finite entanglements}
In this section, we consider the driven dynamics of a linear driving which changes $J$ according to Eqs.~(\ref{gjkt}) and~(\ref{grt}) from the Mott phase to the superfluid phase.

\begin{figure}
	\includegraphics[width=1\columnwidth]{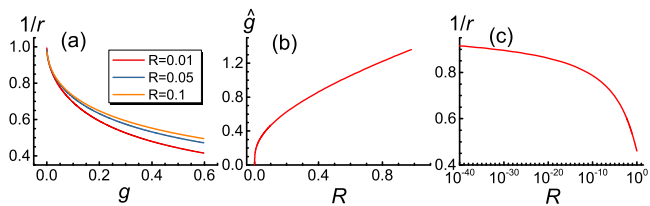}
	\caption{\label{fig:epsart5} (Color online) (a) Inverse rate exponent $1/r$ versus the distance to the critical point $g$ solved from Eq.~(\ref{eq6}) for three driving rates $R$ listed in the legend in the Mott phase. (b) The dependence of the frozen $\hat{g}$ on $R$ from Eq.~(\ref{gw}) for $C\tau_0/\tau(C)=1$. (c) $1/r$ versus $R$ at $g=\hat{g}$ in (b).}
\end{figure}
First of all, as emphasized, a salient feature is that for the KT transition, the rate exponent $r$ depends on both $g$ and $R$ as dictated by Eq.~(\ref{eq6}). In Fig.~\ref{fig:epsart5}(a), we show the dependence of $1/r$ on $g$ for three $R$ values. It is seen that $1/r$ decreases both with increasing $g$ for a fixed $R$ and with $R$ for a fixed $g$. The dependence of $1/r$ on $g$ is similar to the experimental result estimated from ultracold atoms in an optical lattice described by the same BH model~\cite{Braun}. However, the same exponent is estimated from fitting $\xi$ as a power law of $R$ for some $R$ ranges rather than for each $R$~\cite{Braun}. In addition, $1/r$ approaches $1$ only for $g$ very close to $0$, the critical point, in agreement with the experimental results of $1/r<1$~\cite{Braun}. This is understandable since the frozen $g=\hat{g}$ given by Eq.~(\ref{gw}) approaches $0$ only for vanishingly small values of $R$, which is expectable from the steep rise near $R=0$ in Fig.~\ref{fig:epsart5}(b). More explicitly, from the dependence of the inverse rate exponent $1/r$ on $R$ at $g=\hat{g}$ in Fig.~\ref{fig:epsart5}(c), one sees that only extremely small $R$ values can the exponent of unity be acquired in agreement with previous results~\cite{Dzia}. Note that here the exponent $1/r$ is explicitly obtained by solving Eqs.~(\ref{gw}) and~(\ref{eq6}), whereas the previous results are estimated by power-law fits for some ranges of $R$~\cite{Dzia}.

\begin{figure*}
	\includegraphics[width=\textwidth]{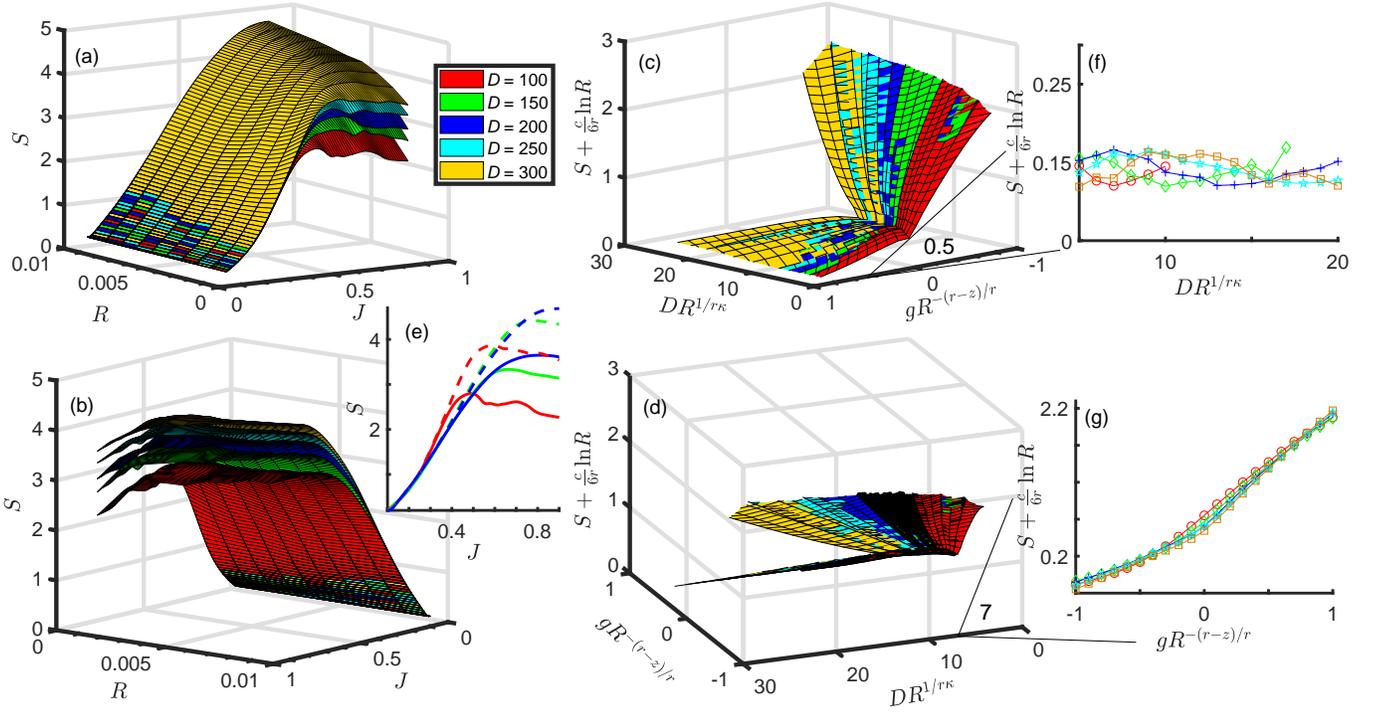}
	\caption{\label{fig:epsart6}(Color online) (a) and (b) Dependence of the entanglement entropy $S$ on $J$ swept linearly at a series of $R$ for five $D$ values colored in the legend in (a), which is shared by all four main panels. (c) and (d) FTS of $S$ according to Eq.~(\ref{eq:rs}). The two rows of the main panels are different only in the angle of view. The inset (e) shows $S$ versus $J$ for $R=0.001$ (red), $R=0.005$ (green), and $R=0.01$ (blue) at $D=100$ (solid) and $D=300$ (dashed lines). The insets (g) and (f) display sections of the main plots at $gR^{-(r-z)/r}=0.5$ and $DR^{1/r\kappa}=7$, respectively. Symbols are 2D interpolations and lines connecting them are only a guide to the eyes and share identical colors with the main panels.}
\end{figure*}
We now present the FTS of the entanglement entropy $S$ in real-time driving. We will not consider effects of phase ordering~\cite{Jelic}, since it has been shown that once the extrinsic self-similarity of the so-called phases fluctuations is considered, critical scaling is good down to quite low temperatures~\cite{Yuan,Yuan1}. In the real-time evolution, the correlation length does not converge and we cannot study its scaling unfortunately. However, the real-time evolution is unitary and the accumulative errors from normalization for the entanglement entropy in imaginary-time does not bother us. In Figs.~\ref{fig:epsart6}(a),~\ref{fig:epsart6}(b) and its inset~\ref{fig:epsart6}(e), we show the variation of $S$ with $J$ for a series of $R$ and $D$. It is seen that as $J$ increases, $S$ reaches a maximum beyond $J_c$, then decreases, and finally oscillates~\cite{Cao}. The maximum tends to move to lower $J$ values for smaller $R$ values at a fixed $D$. This is reasonable because the larger the $R$ value is, the stronger the hysteresis~\cite{Zhong1,Zhong2,Cao}. Also, smaller $R$ values give rise to lower maxima. In addition, a bigger $D$ has larger $S$ values for the same $R$ similar to the imaginary-time evolution. The $S$ surfaces of different $D$ values separate from near $J\simeq J_c$ and all the way to the end of the driving. The large separation of different $D$ surfaces beyond the maxima must mainly be a contribution from the superfluid phase, since it does not appear in the 1D transverse field Ising model~\cite{Cao}. This is in contrast with the behavior in Fig.~\ref{fig:entim} in which $J$ is fixed and $S$ evolves with the imaginary time. The hysteresis due to the driving pushes the Mott phase up to $J>J_c$ to the $S$ maxima, which are representations of the actual transition point~\cite{Cao}. Accordingly, although the scaling theory is designed for the Mott phase which exists for \(J < J_c\) or $g>0$ in equilibrium, we can extend it to $g<0$ to describe the ``superheating" Mott phase in nonequilibrium driving by taking its absolute values as the distance to the critical point again. In Figs.~\ref{fig:epsart6}(c) and~\ref{fig:epsart6}(d), we show the scaled $S$ according to Eq.~(\ref{eq:rs}). It is seen that the surfaces of different bond dimensions splice and merge quite well as exhibited by a mixture of different colors on their overlapping edges. Both $J<J_c$ in the Mott phase and $J>J_c$ for the hysteretic Mott phase within \(-1<gR^{-(r-z)/r}<1\) exhibit good collapses. This is further demonstrated in the two orthogonal sections in Figs.~\ref{fig:epsart6}(f) and~\ref{fig:epsart6}(g). Note that the present 2D interpolations have large errors due to the few data and the small scale in Figs.~\ref{fig:epsart6}(e). Note also that $gR^{-(r-z)/r}$ cannot be simply fixed because $r$ varies with $g$. For $J$ bigger than its value at the $S$ maximum, the system changes to the superfluid phase and the surfaces bend down, collapses are not expected to occur. Therefore, there again exists a finite region that is described by the exponentially growing correlation length in the driving and hence by our scaling theory of FTS once the effect of finite bond dimensions is considered.

\section{Summary\label{summ}}
We have developed a series of scaling theories for the KT phase transitions on the basis of the hallmark exponential growth of their correlation length, Eq.~(\ref{eq1}). The scale transformation of the distance $g$ to the critical point, Eq.~(\ref{eq2}) or~(\ref{eq2x}), has been derived consistently. The resultant dependence on the scale is logarithmic rather than the usual power law and hence may be referred to as logarithmic relevance. Finite-size scaling, Eqs.~(\ref{xiL}) and~(\ref{xiLx}), finite-entanglement scaling, Eq.~(\ref{fes}), short-time critical dynamics, Eqs.~(\ref{eq:four}),~(\ref{eq4}),~(\ref{eq:iment}),~(\ref{xitppx}), and~(\ref{xitppx1}), and finite-time scaling (FTS), Eqs.~(\ref{eq:fts}),~(\ref{eq:ftsr}),~(\ref{eq:rs}),~(\ref{eq:ftst}), and~(\ref{eq:ftstr}), as well as some of their interplay have then been derived accordingly. A quasiequilibrium scaling form at long times and large bond dimensions $D$, Eq.~(\ref{xitpx}), is also given. Finite-size and finite-entanglement scaling forms, Eq.~(\ref{xiLxp}) and a similar one with $D^{\kappa}$ and $f_D$ in place of its $L$ and $f_L$, respectively, somehow similar to but different from a frequently employed ansatz~(\ref{ansatz}) have also be presented, with the finite-entanglement one newly derived. Relaxation times of both a normal power-law, Eq.~(\ref{txiz}), and an anomalous power-law with a logarithmic factor, Eq.~(\ref{teq}), have been considered for the short-time critical dynamics and FTS. The FTS form, along with the driven timescale, can be casted into the usual one, Eq.~(\ref{eq:ftsr}), for a usual critical point. However, the rate exponent $r$, Eq.~(\ref{eq6}), is both $g$ and $R$ dependent for the KT transitions, Fig.~\ref{fig:epsart5}(a). Only when the anomalous relaxation time is considered, both the FTS form, Eq.~(\ref{eq:ftstr}), and the driven timescale, Eq.~(\ref{trrr}), have an additional $R$ factors and hence different from the usual forms. For both kinds of the relaxation time, the frozen conditions that result in the Kibble-Zurek scaling of topological defect density for a linear driving across the KT transition point have been shown to be a particular value of the relevant scaled variable as usual. In particular, at the value, the driven timescale is proportional to the correlation time and the time to the critical point. The exponent for the Kibble-Zurek scaling, Fig.~\ref{fig:epsart5}(c), has been obtained from numerically solving the equations both for the rate exponent and the frozen condition without the need to fit roughly.

To verify the theory, we utilize the KT phase transition of the one-dimensional Bose-Hubbard model at unit filling. The infinite time-evolving-block-decimation (iTEBD) algorithm has been employed to solve numerically the model for the finite bond dimensions $D$. From the imaginary-time evolution of the correlation length $\xi$, we have found that its maximum slope (representing the effective exponent $1/z$) in double logarithmic scales correspond to the straightest part of the curve. This discovery enables us to estimate the KT transition point to be $J_c=0.302(1)$ at the infinite bond dimension. With this critical point, the surfaces of the correlation length evolution for various hopping amplitude $J$ and $D$ collapse well within a certain region according to the scaling theory, indicating that there indeed exists a finite region in which Eq.~(\ref{eq1}), the exponential growth of the correlation length, is valid. We have also shown that the numerical $1/z$ is closer and equal to its theoretical value of $1$ when the scaled variables in the scaling function are fixed as compared with freed. The entanglement entropy in imaginary time is also scaled quite reasonably in consideration of its accumulative errors from normalization. For the real-time driving, we have also shown that the FTS with finite bond dimensions describes well the scaling collapse of the entanglement entropy for a series of the driving rate $R$ and $D$ both for $J<J_c$ in the usual Mott phase and for $J>J_c$ in the hysteretic Mott phase. These numerical results therefore confirm the scaling theories for the KT phase transitions.

We note that the scaling theories, built on Eq.~(\ref{eq1}) alone, need to be modified when corrections to the exponential growth of the correlation length matter. Conversely, when the scaling described by the theories is poor, it may well indicate that such corrections have to be taken into account.

\begin{acknowledgments}
We would like to thank Shaolong Zeng for his discussions, Gary A. Williams for his instructive communications, and Rui-Zhen Huang for his code support. This work was supported by the National Natural Science Foundation of China (NNSFC) (Grant No. 11575297). FZ is additionally supported by NNSFC (Grant No. 12175316), while SY was supported by NNSFC (Grant No. 12075324).
\end{acknowledgments}



\end{document}